\documentclass[12pt]{iopart}

\expandafter\let\csname equation*\endcsname\relax
\expandafter\let\csname endequation*\endcsname\relax
\usepackage{amsmath, amssymb, setspace}

\usepackage{cleveref}
\usepackage{etoolbox}

\crefname{equation}{}{}

\newcommand{\mathsym}[1]{{}}
\newcommand{\unicode}[1]{{}}

\patchcmd{\numparts}{\addtocounter{equation}{1}}{\refstepcounter{equation}}{}{}

\usepackage{graphicx}
\usepackage{epstopdf}
\usepackage{algorithmic}
\ifpdf
  \DeclareGraphicsExtensions{.eps,.pdf,.png,.jpg}
\else
  \DeclareGraphicsExtensions{.eps}
\fi
\usepackage[caption=false]{subfig}
\usepackage{verbatim}
\usepackage{xcolor}

\newcommand{\del}{\partial}

\usepackage{ulem}

\begin{document}


\title[]{Magnetic response of topological insulator layer with metamaterial substrate induced by an electric point source}

\author{Qiang Sun${}^{1,*}$, Eitan Dvorquez${}^2$, Felipe Pinto${}^2$, Mohan C. Mathpal${}^2$, Jerónimo R. Maze${}^2$, Brant C. Gibson${}^{1}$, Andrew D. Greentree${}^{1}$}

\address{${}^1$Australian Research Council Centre of Excellence for Nanoscale Biophotonics, School of Science, RMIT University, Melbourne, VIC 3001,
Australia.\\
${}^2$Institute of Physics, Pontificia Universidad Católica de Chile, Santiago, Chile.}
\ead{qiang.sun@rmit.edu.au}
\vspace{10pt}
\begin{indented}
\item[]July 2024
\end{indented}

\begin{abstract}
Topological insulators (TIs) are materials with unique surface conductive properties that distinguish them from normal insulators and have attracted significant interest due to their potential applications in electronics and spintronics. However, their weak magnetic field response in traditional setups has limited their practical applications. Here, we show that integrating TIs with active metamaterial substrates can significantly enhance the induced magnetic field by more than $10^4$ times. Our results demonstrate that selecting specific permittivity and permeability values for the active metamaterial substrate optimizes the magnetic field at the interface between the TI layer and the metamaterial, extending it into free space. This represents a substantial improvement over previous methods, where the magnetic field decayed rapidly. The findings reveal that the TI-metamaterial approach enhances the magnetic field response, unveiling new aspects of TI electromagnetic behavior and suggesting novel pathways for developing materials with tailored electromagnetic properties. The integration of metamaterials with TIs offers promising opportunities for advancements in materials science and various technological applications. Overall, our study provides a practical and effective approach to exploring the unique magnetic field responses of TIs, potentially benefiting other complex material systems.
\end{abstract}

\noindent{\it Keywords\/}: Topological Insulator; Metamaterial; Magnetic Field Enhancement; Hybrid Layer Structure

%
%
%
%
%

\section{Introduction}

The study of topological insulators (TIs), materials with a unique topological order, represents a significant step forward in developing new electronics~\cite{Anirban2023}. These materials are insulators internally but have conductive surfaces~\cite{Hasan2010, Qi2011, Bernevig2013-ia, Ortmann2015-iq, Mogi2017}. The ability to exhibit conductive edge states without an external magnetic field, as demonstrated in the quantum spin Hall effect in two-dimensional TIs, highlights their potential for transforming electronic devices~\cite{Toriyama2025}.

The special properties of TIs have created novel applications that are not possible with conventional insulators. Qi {\it et al}.~\cite{Qi2009} demonstrated that an electric charge near a TI surface induces a magnetic monopole due to the topological magneto-electric effect. Martín-Ruiz {\it et al}.~\cite{MartnRuiz2015, MartnRuiz2016} used the Green's function method to study boundary effects and special electromagnetic properties in TIs with Chern-Simons extended electrodynamics. The research by Lai {\it et al}.~\cite{Lai2014} explored the tunability of plasmons in TIs, showing their potential in plasmonics and spintronics. Wang {\it et al}.~\cite{Wang2021} discussed intrinsic magnetic TIs, such as MnBi$_2$Te$_4$, which combine magnetism with topological properties, leading to new quantum phenomena. 

Meanwhile, synthetic TI structures~\cite{Tian2017, Yue2018}, such as nanowires and nanoribbons, are being explored in field-effect transistors (FETs), optoelectronic devices, memory storage devices and magnetoelectric devices~\cite{Legg2022, Fischer2022, Mnning2021, CastroEnrquez2022, Wu2021}. These structures exhibit enhanced surface conduction due to their high surface-to-volume ratio and can improve the performance of devices like photodetectors, spintronic devices, and quantum computing systems~\cite{He2019, Pandey2021, Breunig2021} by leveraging their unique surface properties and quantum effects.

The novel magnetic field response of TIs is an area of growing interest due to its potential applications in advanced technology. The unique topological electromagnetic properties of TIs are key to understanding their novel magnetic field responses, although these effects are usually small. To enhance their electronic and magnetic properties, the development of layered structures incorporating TIs has been proposed~\cite{Dvorquez2024}. In 2011, Burkov and Balents~\cite{Burkov2011} proposed Weyl semimetal states in TI multilayers to achieve unusual electronic states. Further studies explored TI-ferromagnet multilayers~\cite{Aftab2020} for efficient spin-orbit torque~\cite{Ghosh2018, Fan2022}, highlighting practical applications in spintronic devices~\cite{Devlin2021}. In 2020, An {\it et al}.~\cite{An2020} demonstrated significant polarization rotation in multilayer TI structures, enhancing optical applications. Additionally, Ardakani and Zare (2021)~\cite{GhasempourArdakani2021} showed strong Faraday rotation using dielectric multilayers with a single TI layer with improved magnetic field interactions. 

Most of the work reviewed above focuses on the dynamic effects. In this work, we explore the novel magnetic field response of TIs when exposed to a static electric point source. This static response, which couples electric and magnetic fields, distinguishes TIs from normal insulators more significantly than their dynamic behavior. However, as demonstrated by Qi {\it et al}.~\cite{Qi2009}, the induced magnetic field by a TI is very weak, making it challenging to harness for practical applications. To address this challenge, we explore the magnetic field response of TIs in composite layered structures, particularly with metamaterials.

Metamaterials have been a significant area of research since the late 20th century, with their unique properties offering potential for various technological applications. The concept was first theoretically proposed by Victor Veselago in 1968~\cite{Veselago1968}, who suggested that materials could exhibit both negative permittivity and permeability, resulting in a negative refractive index. This idea, though intriguing, remained largely theoretical until the development of practical metamaterials in the early 2000s~\cite{Ozbay2007}. In 2000, Smith and Kroll~\cite{Smith2000} demonstrated negative refraction in left-handed materials, a groundbreaking experimental validation that paved the way for further research and development. Following this, Shelby {\it et al}.~\cite{Shelby2001} provided experimental verification of a negative index of refraction using a metamaterial constructed from an array of copper split-ring resonators and wires. Liu {\it et al}.~\cite{Liu2016} introduced dynamic and tunable metamaterials that can be electrically controlled, using vanadium dioxide to achieve multifunctional control. This marked a significant advancement, enabling the modulation of metamaterial properties through external stimuli, thereby enhancing their practical applicability. Most recently, Castles {\it et al}. in 2020~\cite{Castles2020} explored active metamaterials with static electric susceptibility less than one. This work demonstrated that active metamaterials could manipulate static electric susceptibility, a property that had been theoretically suggested but never experimentally confirmed until then.

In this work, we focus on the magnetic response of TIs by integrating them with multi-layer composite structures with metamaterials. By employing the unique electromagnetic properties of TIs and metamaterials, we investigate the magnetic response induced by an electric charge. This study explores a novel approach to enhancing the magnetic field response of TIs, highlighting unique electromagnetic interactions within TIs and opening new avenues for their practical application in advanced technologies.

\section{Model}\label{sec:model}

For a standard dielectric material, we have $\bi{D} = (\epsilon_{0} \epsilon_{r}) \bi{E}$ for electric displacement, $\bi{D}$, and electric field, $\bi{E}$, with $\epsilon_{0}$ and $ \epsilon_{r}$ being the vacuum permittivity and relative permittivity of the material, respectively; and $\bi{H} = \bi{B}/(\mu_{0}\mu_{r})$ for magnetic induction strength, $\bi{B}$ and field, $\bi{H}$, with $\mu_{0}$ and $\mu_{r} $ being the vacuum permeability and relative permeability of the material, respectively. These equations illustrate that, typically, the electric and magnetic fields in a dielectric material are independent of each other. However, this independence does not apply to chiral materials, such as TIs, which exhibit a coupling between electric and magnetic fields. Specifically, for TIs, the constitutive equations become:
\numparts
\label{eq:ConseqTI0}
\begin{eqnarray}  
\bi{D} &= \epsilon_{0} \epsilon_{r} \bi{E} - \frac{\alpha \theta}{\pi} \epsilon_{0} c \bi{B}, \\
\bi{H} &= \frac{\bi{B}}{\mu_{0}\mu_{r}} + \frac{\alpha \theta}{\pi} \frac{1}{\mu_0 c} \bi{E} ,
\end{eqnarray}
\endnumparts 
where $\theta$ is the topological phase of the TI, $c$ is the speed of light and $\alpha$ is the fine structure constant.

\begin{figure}[t]
\centering{}
\includegraphics[width=0.6\textwidth]{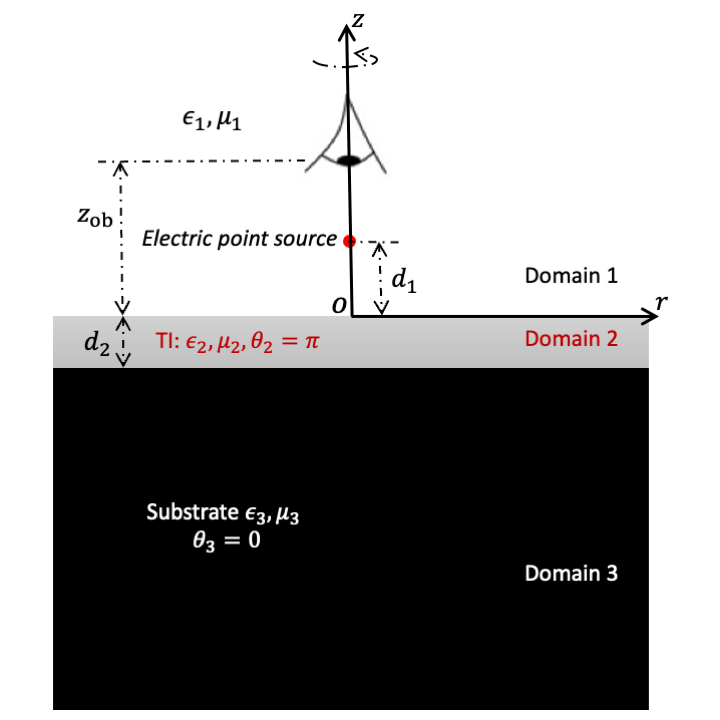}
\caption{Sketch of induced magnetic monopole of TI from an electric point charge by multi-layer composite composed with TI, normal insulators, $\mu$-metals.}  \label{Fig:SQSketch}
\end{figure}

In scenarios where the TI has a substantial thickness (beyond a few dozens of nanometers), the TI phase, $\theta$, stabilises as a constant throughout the material, as illustrated in Fig.~\ref{Fig:SQSketch}. As such, we have $\nabla \theta = \mathbf{0}$ within the TI, while phase transitions are confined to thin layers (less than a few nanometers) at the interfaces between the TI and its surrounding environment, which indicates that these transitions can be accounted for by adjusting boundary conditions. Considering the setup displayed in Fig.~\ref{Fig:SQSketch}, without any free charges the macroscopic Maxwell's equations are $\nabla \cdot \bi{D} = 0$ and $\nabla \cdot \bi{B} = 0$. As such, from Eq.~(\ref{eq:ConseqTI0}), we obtain
\label{eq:ConseqTI2}
\begin{eqnarray}
\bi{\nabla} \bi{\cdot} \bi{E} &= 0, \qquad
\bi{\nabla} \bi{\cdot} \bi{B} &= 0. 
\end{eqnarray}


Given these conditions, for static problems, the electric and magnetic fields can be expressed in terms of electric ($\phi_{E}$) and magnetic ($\phi_{B}$) potentials as
\begin{eqnarray}\label{eq:EBphi}
\bi{E} = - \nabla \phi_{E}, \qquad \bi{B} = -\nabla \phi_{B}.
\end{eqnarray}
Both potentials satisfy the Laplace equation, as
\begin{eqnarray}\label{eq:LaplaceEq}
\nabla^2 \phi = 0
\end{eqnarray}
with $\phi$ representing either $\phi_{E}$ or $\phi_{B}$.

In the context of a two-layer structure as depicted in Fig.~\ref{Fig:SQSketch}, where domain 2 is a TI layer with thickness of $d_2$, and domain 3 represents an infinitely thick structure, we consider the free space above the TI layer as domain 1 with relative permittivity $\epsilon_1=1$ and relative permeability $\mu_{1}=1$. A static electric point charge $Q$ is located $d_1$ above the top surface of the TI layer (domain 2). Our model generalises the two layers as TIs, each characterised by specific material properties: $\epsilon_{i}$ the relative permittivity, $\mu_{i}$ the relative permeability  and $\theta_{i}$ the topological phase of the material where $i=2,3$ indexes the domain. This framework allows for the simplification of a normal insulator as a TI with a phase $\theta=0$.

Considering the two-layer structure described above, we explore the impact of thickness and material properties on the induced fields due to the presence of TI. The analysis is then simplified by assuming cylindrical symmetry in which the size of the TI layer along its radial direction is much larger than its thickness. We employed the zeroth-order Hankel transform~\cite{Poularikas2009-dt, Rahman2017} to solve our model. It is worth noting that for the cases with a single interface as shown in Ref.~\cite{Qi2009}, the image charge method works well to describe the induced charge distribution. However, when multiple interfaces are involved—even just two—the image method becomes extremely tedious, requiring the consideration of an infinite series of image charges, in particular, when considering the coupled electric and magnetic fields with TIs. We then chose to employ the Hankel transform, which provides an efficient and transparent solution to the problem and clearly captures the physics we are exploring. 

Given the axisymmetric nature of the problem, it is practical to employ a cylindrical coordinate system $(r,z)$. Thus, Eq.~(\ref{eq:LaplaceEq}) can be rewritten as
\begin{eqnarray}\label{eq:LaplaceEqCyl}
\frac{1}{r} \frac{\del}{\del r}\left[r \frac{\del \phi(r,z)}{\del r}\right] + \frac{\del^2 \phi(r,z)}{\del z^2} = 0.
\end{eqnarray}
This symmetry simplifies the Laplace equation for electric and magnetic potentials to a form that can be efficiently solved using the zeroth-order Hankel transform, since it allows us to reduce the problem to a set of ordinary differential equations in the vertical direction ($z$) with just a single integral in the radial direction ($r$). This separation of variables leads to a tractable and transparent solution method. Though it is still possible to perform $n$th order Hankel transforms to our model ($n>0$), it will introduce additional integrals involving special functions, as detailed in Chapter 9 of Ref.~\cite{Poularikas2009-dt}. This would make the analysis significantly more cumbersome without providing further physical insight. 

When applying the zeroth-order Hankel transform to the potential, we have
\begin{eqnarray}\label{eq:HankelTrans}
{\mathrm{H}}_{0}\left[\phi(r,z); \rho \right] \equiv \Phi(\rho,z) = \int_{0}^{\infty} \phi(r,z) J_{0}(\rho r) r \,\rmd r,
\end{eqnarray}
and 
\begin{eqnarray}\label{eq:insHankelTrans}
\phi(r,z) = \int_{0}^{\infty} \Phi(\rho,z) J_{0}(\rho r) \rho \,\rmd \rho,
\end{eqnarray}
in which $J_{0}(\rho r)$ is the zeroth-order Bessel function of the first kind. By using the property
\begin{eqnarray}
{\mathrm{H}}_{0}\Bigg\{\frac{1}{r} \frac{\del}{\del r}\left[r \frac{\del \phi(r,z)}{\del r}\right]; \rho \Bigg\} = -\rho^2 {\mathrm{H}}_{0}\left[\phi(r,z); \rho \right],
\end{eqnarray}
Eq.~(\ref{eq:LaplaceEqCyl}) is transformed into
\begin{eqnarray}\label{eq:Hankel0Eq}
\frac{\del^2 \Phi(\rho,z)}{\del z^2} - \rho^2 \Phi(\rho,z) = 0.
\end{eqnarray}
This transform efficiently handles the radial component, reducing the problem in Eq.~(\ref{eq:LaplaceEqCyl}) to one that is more straightforwardly solvable, for instance, this method simplifies the mathematical treatment by focusing on the transform domain, where the solutions to the Laplace equation exhibit exponential behaviours, as $\exp{[\pm \rho z]}/\rho$. As such, in domain 1, we have
\numparts
\label{eq:Phi1}
\begin{eqnarray}
\Phi_{E}^{1} &= C_{1} \frac{\exp{[- \rho z]}}{\rho}, \\
\Phi_{B}^{1} &= C_{2} \frac{\exp{[- \rho z]}}{\rho}.
\end{eqnarray}
\endnumparts
Similarly, in domain 2, we have
\numparts
\label{eq:Phi2}
\begin{eqnarray}
\Phi_{E}^{2} &= C_{3} \frac{\exp{[\rho z]}}{\rho} + C_{5} \frac{\exp{[-\rho (z+d_{2})]}}{\rho }, \\
\Phi_{B}^{2} &= C_{4} \frac{\exp{[\rho z]}}{\rho} + C_{6} \frac{\exp{[-\rho (z+d_{2})]}}{\rho };
\end{eqnarray}
\endnumparts
and in domain 3, we have
\numparts
\label{eq:Phi3}
\begin{eqnarray}
\Phi_{E}^{3} &= C_{7} \frac{\exp{\rho (z+d_{2})]}}{\rho}, \\
\Phi_{B}^{3} &= C_{8} \frac{\exp{\rho (z+d_{2})]}}{\rho}.
\end{eqnarray}
\endnumparts
In Eqs.~(\ref{eq:Phi1}) to (\ref{eq:Phi3}), $C_{1}, ..., C_{8}$ are unknown coefficients to be determined by the boundary conditions at each interface.

The potential due to an electric point source $Q$ located at $(0,\,0,\,d_1)$ in domain 1 is expressed as
\begin{eqnarray}\label{eq:pointsource_rz}
\phi^{s}_{E}(r,z) = \frac{Q}{4\pi \epsilon_{0}\epsilon_{1}} \frac{1}{\sqrt{r^2 + (z-d_1)^2}}.
\end{eqnarray}
Applying the Hankel transform in Eq.~(\ref{eq:HankelTrans}) to Eq.~(\ref{eq:pointsource_rz}), we derived its expression in the transformed space as
\begin{eqnarray}
\fl \Phi^{s}_{E}(q,z) = \frac{Q}{4\pi \epsilon_{0}\epsilon_{1}}  \int_{0}^{\infty} \frac{1}{\sqrt{r^2 + (z-d_1)^2}} J_{0}(\rho r) r \,\rmd r = \frac{Q}{4\pi \epsilon_{0}\epsilon_{1}}  \frac{\exp{(-\rho |z-d_{1}|)}}{\rho}.
\end{eqnarray}

The boundary conditions at the interfaces between domains ensure continuity of the electric and magnetic components across these boundaries. Specifically, they enforce the matching of radial electric and magnetic fields ($E_{r}$, $H_r$) and the normal components of electric and magnetic displacements ($D_{z}$, $B_{z}$) across interfaces, leveraging the transformed potentials. By using the expressions in Eqs.~(\ref{eq:Phi1}) to (\ref{eq:Phi3}) and the relationship in Eq.~(\ref{eq:ConseqTI0}), we have, on interface 1-2 at $z=0$,
\numparts 
\label{eq:C1C8_12}
\begin{eqnarray} 
\fl C_{1} - C_{3} - \exp{[-\rho d_{2}]} \, C_{5}  = -\frac{Q}{4\pi \epsilon_{0}\epsilon_{1}} \exp{[-\rho d_{1}]}, \\
\fl \frac{1}{\mu_1} C_{2}  - \frac{1}{\mu_{2}} \Big\{ C_{4} + \exp{[-\rho d_{2}]} \, C_{6}  \Big\} -  \frac{\theta_{2} \alpha}{\pi } \frac{1}{c} \Big\{ C_{3} + \exp{[-\rho d_{2}]} \, C_{5} \Big\} =0, \\
\fl \epsilon_{1}C_{1} +\epsilon_{2} \Big\{ C_{3} - \exp{[-\rho d_{2}]} \, C_{5} \Big\} - \frac{\theta_{2} \alpha}{\pi } c \Big\{ C_{4} - \exp{[-\rho d_{2}]} \,  C_{6}\Big\} =\frac{Q}{4\pi \epsilon_{0}} \exp{[-\rho d_{1}]}, \\
\fl C_{2}  +  C_{4} - \exp{[-\rho d_{2}]} \, C_{6} =0.
\end{eqnarray}
\endnumparts
On interface 2-3 at $z=-d_{2}$, we can write
\numparts
\label{eq:C1C8_23}
\begin{eqnarray}  
\fl \exp{[-\rho d_{2}]} \, C_{3} + C_{5} - C_{7} = 0, \\
\fl \frac{1}{\mu_{2}} \Big\{\exp{[-\rho d_{2}]} \,  C_{4} + C_{6}  \Big\} + \frac{\theta_{2} \alpha}{\pi } \frac{1}{c} \Big\{\exp{[-\rho d_{2}]} \,  C_{3} + C_{5} \Big\}  - \frac{1}{\mu_{3}} C_{8}  = 0,\\ 
\fl -\epsilon_{2} \Big\{\exp{[-\rho d_{2}]} \,  C_{3} - C_{5} \Big\} + \frac{\theta_{2} \alpha}{\pi } c \Big\{\exp{[-\rho d_{2}]} \, C_{4} - C_{6}\Big\} + \epsilon_{3}C_{7}   = 0,\\
\fl -\exp{[-\rho d_{2}]} \, C_{4} + C_{6} + C_{8} =0.
\end{eqnarray}
\endnumparts
Solving the linear system formed by these boundary conditions in Eqs.~(\ref{eq:C1C8_12}) and (\ref{eq:C1C8_23}) yields the unknown coefficients, $C_{1}, ..., C_{8}$. Thereafter, we can determine the potential distributions in each domain, which, upon applying the inverse Hankel transform as in Eq.~(\ref{eq:insHankelTrans}), yield the spatial distributions of electric and magnetic fields via Eq.~(\ref{eq:EBphi}). This comprehensive approach, grounded in boundary matching and transforms, enables a detailed exploration of the electromagnetic response of TIs to external electric charges, through which we will illuminate the effects of thickness and material properties on induced fields within these advanced materials in the next Section.

\section{Results}\label{sec:results}
Let us first revisit the simple scenario where a single electric point source is positioned near a large TI \cite{Qi2009}. In this setup, the space is divided into two regions: the upper half is air and the lower half is the TI material. The electric field ($\bi{E}$) and magnetic field ($\bi{B}$) expressions in the air domain are given by
\numparts
\label{eq:EBTIhalf}
\begin{eqnarray}
\fl \bi{E}(\bi{x}) = \frac{Q}{4\pi \epsilon_{0}\epsilon_{1}} \left[ \frac{\bi{x}-\bi{x}_{0}}{|\bi{x}-\bi{x}_{0}|^3} -\frac{(\epsilon_{2}/\epsilon_{1}-1)(\mu_{1}/\mu_{2}+1)+\bar{\theta}^2}{(\epsilon_{2}/\epsilon_{1}+1)(\mu_{1}/\mu_{2}+1)+\bar{\theta}^2} \frac{\bi{x}+\bi{x}_{0}}{|\bi{x}+\bi{x}_{0}|^3}  \right],\\
\fl \bi{B}(\bi{x}) =-\frac{Q}{4\pi \epsilon_{0}\epsilon_{1} c} \frac{2\bar{\theta}}{(\epsilon_{2}/\epsilon_{1}+1)(\mu_{1}/\mu_{2}+1)+\bar{\theta}^2} \frac{\bi{x}+\bi{x}_{0}}{|\bi{x}+\bi{x}_{0}|^3} 
\end{eqnarray}
\endnumparts
in which $\bi{x}$ is the observation location, $\bi{x}_0$ the location of the point source, and $\bar{\theta} \equiv (\alpha \theta)/\pi$. This setup reflects the conditions sketched in Fig.~\ref{Fig:SQSketch} with the TI layer and its substrate having identical properties, i.e. $\epsilon_2 = \epsilon_3$, $\mu_2 = \mu_3$, $\theta_2 = \theta_3$ and with the point source located at $\bi{x}_0=(0,0,d_1)$ and the observation point at $\bi{x} = (0,0,z_{\mathrm{ob}})$. This configuration can not only be used to validate our model but also highlights the typically weak magnetic response induced by an electric point source near a TI. Additionally, it suggests methods to enhance this response.

The small magnitude of the magnetic field induced by the TI, evident from Eq.~(\ref{eq:EBTIhalf}b), is primarily due to the field’s amplitude being linearly proportional to the charge of the point source but inversely proportional to the speed of light and to the square of the observation distance to the TI surface. For instance, an electric point source carrying 200 elementary charges, positioned at $d_1 = 2 $ $ \mu\mathrm{ m}$ from the TI surface when $\epsilon_2 = 20$ and $\mu_2 = 1$, would produce a magnetic field at an observation point at $z_{\mathrm{ob}} = 600 \mathrm{ }\mu$m away with an amplitude of $9.211 \times 10^{-13} $ T (0.9211 pT) according to Eq.~(\ref{eq:EBTIhalf}b). For the same case calculated by our model, as demonstrated in Sec. 2, good agreement has been found to the result obtained by Eq.~(\ref{eq:EBTIhalf}b) with a difference of less than $2.50 \times 10^{-25}$ T, which is within the numerical accuracy of the computer. This highlights the accuracy of our model and calculations. 

To enhance the magnetic field, one potential way is to minimize the denominator of the expression in Eq.~(\ref{eq:EBTIhalf}), in particular, to make:
\begin{eqnarray}
\left(\frac{\epsilon_{2}}{\epsilon_{1}}+1\right)\left(\frac{\mu_{1}}{\mu_{2}}+1\right)+\bar{\theta}^2 \rightarrow 0.
\end{eqnarray}
To achieve this reduction, this would require that the material property term $(\epsilon_{2}/\epsilon_{1}+1)(\mu_{1}/\mu_{2}+1)$ approach negative values, which is unfeasible with natural materials since their relative permittivity and permeability are positive. Adjusting the properties of TIs while maintaining their topological characteristics is challenging. However, with recent advancements in synthetic metamaterials that exhibit negative permittivity and permeability, it is possible to use these materials together with a TI to form layered structures, as depicted in Fig.~\ref{Fig:SQSketch}, to enhance the magnetic response of TIs to electric point sources. 

As illustrated by the solution procedure at the end of Sec~\ref{sec:model}, and particularly based on the relationships in Eqs. (\ref{eq:insHankelTrans}) and (\ref{eq:EBphi}), we can conclude that if certain material property parameters cause the denominator of coefficient $C_2$ for $\Phi_B^1$ to approach zero, the magnetic field in the free space (domain 1) will be significantly enhanced. After performing algebraic manipulation to solve Eqs.~(\ref{eq:C1C8_12}) and (\ref{eq:C1C8_23}), we found the expression of $C_2$ and its denominator $I_{dn}$, as detailed in~\ref{sec:A}. It is shown that $I_{dn}$ depends on both $\epsilon_3$ and $\mu_3$ as well as on the properties of the TI layer. Additionally, $I_{dn}$ is also a function of the Hankel transform variable $\rho$. In Fig.~\ref{Fig:B_denom}, with respect to different values of $\rho$, the variations of $I_{dn}$ are shown when the substrate relative permittivity $\epsilon_3$ and permeability $\mu_3$ sweep from -40 to 40, with the properties of the TI layer being $d_2 = 600 $ $ \mu\mathrm{ m}$, permeability $\mu_2=1$, relative permittivity $\epsilon_2$=20 and phase $\theta_2 = \pi$. It is evident that several combinations of $\epsilon_3$ and $\mu_3$ can cause $I_{dn}$ to approach zero for different values of $\rho$, in particular when either $\epsilon_3$ or $\mu_3$ or both are negative. This suggests that it is possible to enhance the magnetic field induced by a TI driven by a point source with a metamaterial substrate.

\begin{figure}[!t]
\centering{}
\subfloat[]{ \includegraphics[width=0.45\textwidth]{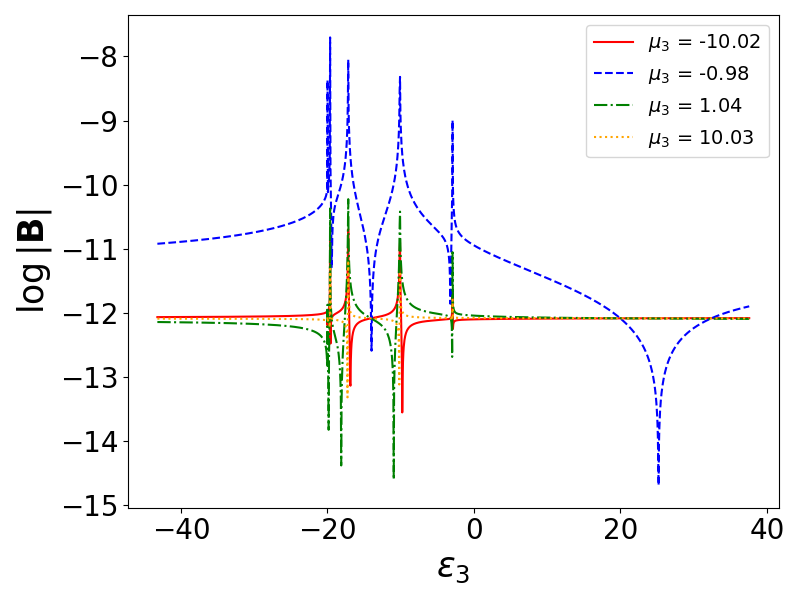}}  \quad
\subfloat[]{ \includegraphics[width=0.45\textwidth]{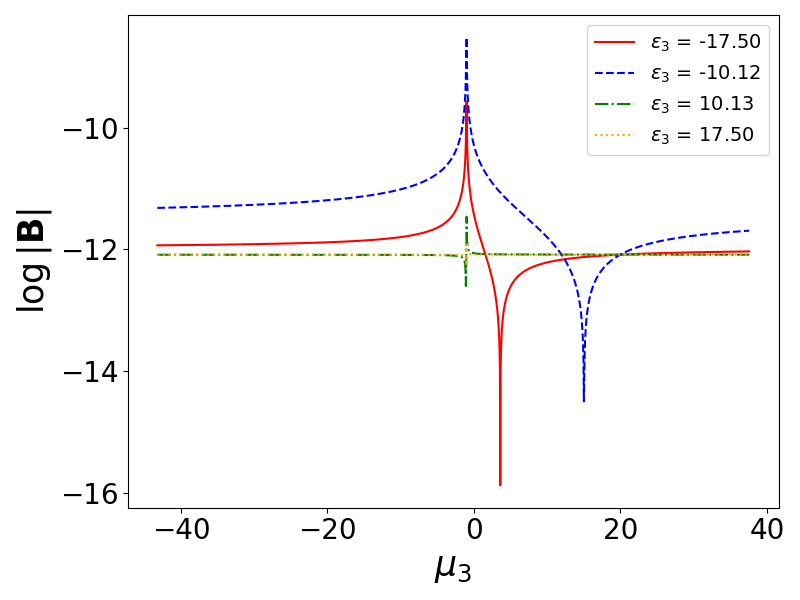}} 
\caption{The logarithm (base 10) of magnetic field strength, $\log{|\bi{B}|}$ at $z_{\mathrm{ob}} = 600 $ $ \mu\mathrm{ m}$, generated by a TI when interfaced due to a 200-elementary-charge point electric source at $d_1 = 2$ $ \mu\mathrm{ m}$ with various substrates when its relative permittivity (a) $\epsilon_3$ and (b) permeability $\mu_3$ sweep from -35 to 35 while maintaining a constant thickness $d_2 = 600 $ $ \mu\mathrm{ m}$, $\epsilon_2=20$, $\mu_2=1$ and $\theta_2 = \pi$. The bright peak features indicate conditions for maximum magnetic field enhancement, serving as a crucial guide for substrate selection aimed at boosting magnetic responses in TI applications. } \label{Fig:mu2_1lines}
\end{figure}
\begin{figure}[!t]
\centering{}
\subfloat[$\epsilon_{2} = 10$, $\mu_{2} = 1$]{ \includegraphics[width=0.45\textwidth]{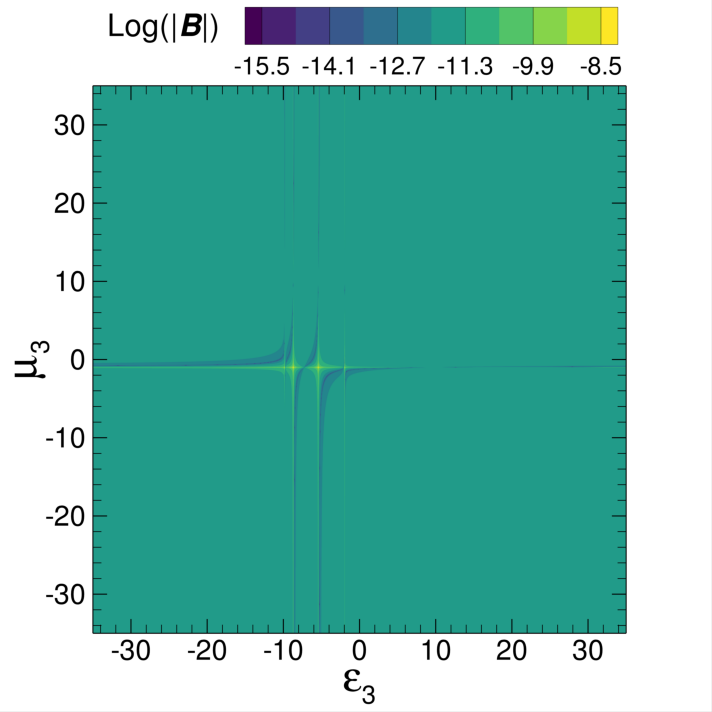}}  \quad
\subfloat[$\epsilon_{2} = 15$, $\mu_{2} = 1$]{ \includegraphics[width=0.45\textwidth]{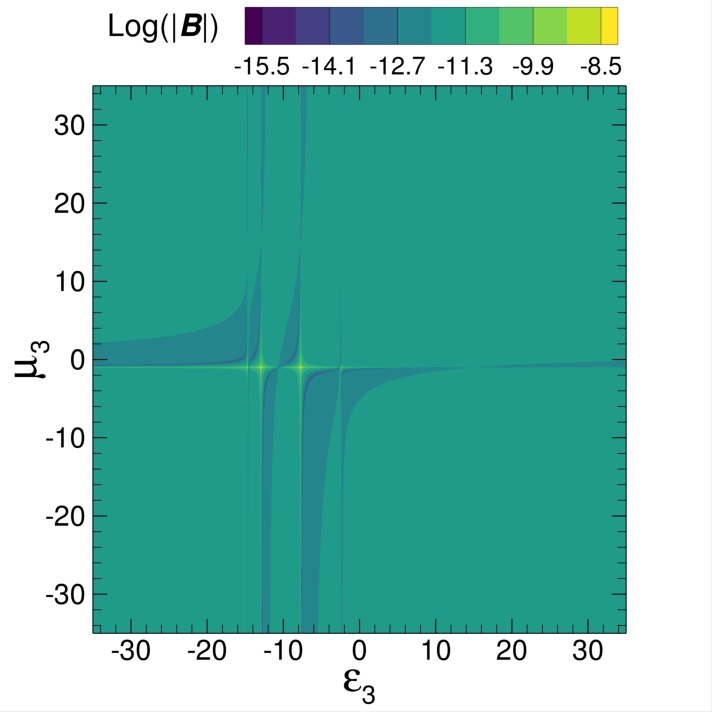}} \\
\subfloat[$\epsilon_{2} = 20$, $\mu_{2} = 1$]{ \includegraphics[width=0.45\textwidth]{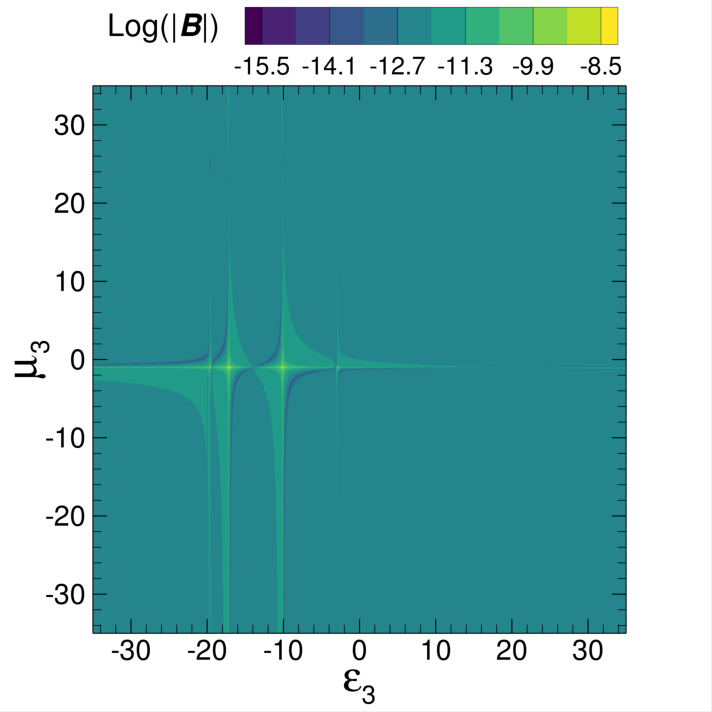}}  \quad
\subfloat[$\epsilon_{2} = 25$, $\mu_{2} = 1$]{ \includegraphics[width=0.45\textwidth]{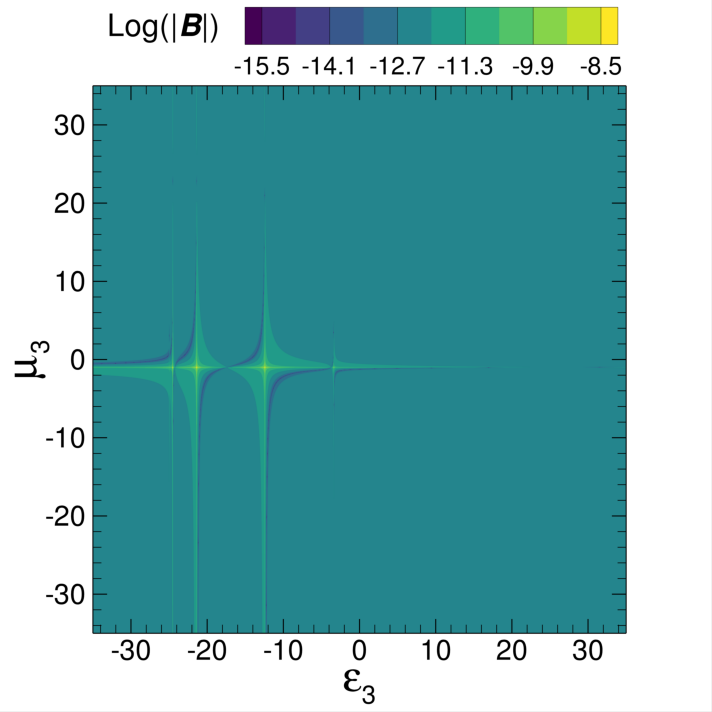}}
\caption{The logarithm (base 10) of magnetic field strength, $\log{|\bi{B}|}$ at $z_{\mathrm{ob}} = 600 $ $ \mu\mathrm{ m}$, generated by a TI when interfaced due to a 200-elementary-charge point electric source at $d_1 = 2$ $ \mu\mathrm{ m}$ with various substrates when its relative permittivity $\epsilon_3$ and permeability $\mu_3$ sweep from -35 to 35. Each sub-figure corresponds to a distinct relative permittivity $\epsilon_2$ of the TI layer: (a) $\epsilon_2$=10, (b) $\epsilon_2$=15, (c) $\epsilon_2$=20, and (d) $\epsilon_2$=25, while maintaining a constant thickness $d_2 = 600 $ $ \mu\mathrm{ m}$, permeability $\mu_2=1$ and $\theta_2 = \pi$. The bright peak features indicate conditions for maximum magnetic field enhancement, serving as a crucial guide for substrate selection aimed at boosting magnetic responses in TI applications. The sharpness and position of these peaks shift to more negative relative permittivity of the metamaterial substrate with the increase of $\epsilon_2$.} \label{Fig:mu2_1}
\end{figure}

To study the magnetic response of a TI-metamaterial hybrid structure, we explore the substrate properties when the substrate varies between a standard insulator and a metamaterial with no topological effects ($\theta_3 = 0$). The TI layer was consistently configured with a phase of $\theta_2 = \pi$ and a thickness of $d_2 = 600  $ $ \mu\mathrm{ m}$. 

Fig.~\ref{Fig:mu2_1lines} illustrates the logarithmic magnitude of the induced magnetic field in free space at $z_{\text{ob}}=600\,\mu\text{m}$ as a function of the substrate’s relative permittivity $\epsilon_3$ and permeability $\mu_3$ when the TI's $\epsilon_2=20$ and $\mu_2=1$. The contour maps in this figure reveal distinct resonant regions—manifested as bright peaks—where the denominator $I_{\text{dn}}$ in the expression for $C_2$ as shown in~\ref{sec:A} approaches zero, leading to significant magnetic field enhancement. Notably, these resonant enhancements occur primarily when either $\epsilon_3$ or $\mu_3$, or both, assume negative values, a behavior characteristic of active metamaterials.

We then further systematically explore the TI's relative permittivity, $\epsilon_2$, with four values: 10, 15, 20, and 25. The relative permittivity, $\epsilon_3$ and permeability, $\mu_3$ of the metamaterial substrates were explored over a range from -35 to 35 with a step of 0.05. Our results, which are illustrated in Fig.~\ref{Fig:mu2_1}, highlight significant variations in the magnetic field's amplitude in response to changes in the substrate properties. 

Fig.~\ref{Fig:mu2_1} displays the logarithmic magnitude of the magnetic field, $\log{|\bi{B}|}$, at $z_{\mathrm{ob}} = 600  $ $ \mu\mathrm{ m}$ generated by an electric point source with 200 elementary charges at $d_1 = 2 $ $ \mu\mathrm{ m}$ as a function of $\epsilon_3$ and $\mu_3$ of the substrate. These plots indicate possible regions where the magnetic field can be significantly enhanced, appearing as pronounced peaks against a background of relatively low magnetic field intensity. From Fig.~\ref{Fig:mu2_1} (a) to (d), we observe that as the relative permittivity, $\epsilon_2$, of the TI layer increases from 10 to 25, the regions of peak enhancement shift to more negative relative permittivity of the metamaterial substrate, suggesting a complex interplay between the TI's permittivity and the substrate's properties. Particularly striking are the sharp resonant-like features that indicate conditions under which the magnetic response is markedly amplified.

\begin{figure}[t]
\centering{}
\subfloat[]{ \includegraphics[width=0.45\textwidth]{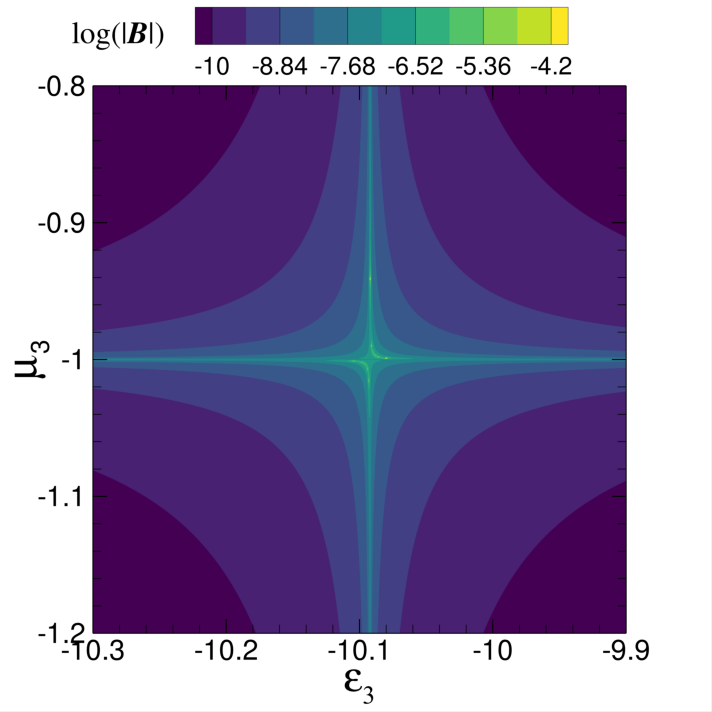}}  \quad
\subfloat[]{ \includegraphics[width=0.45\textwidth]{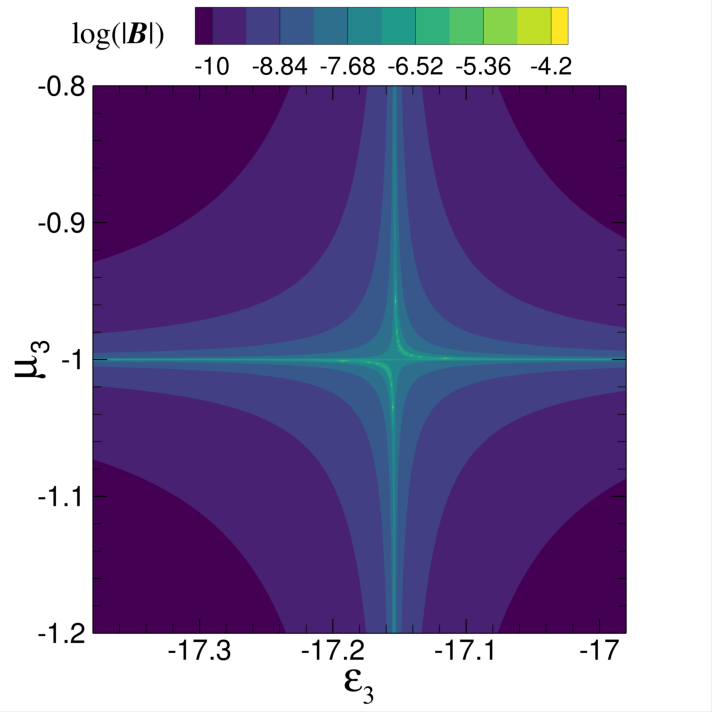}} 
\caption{The logarithm (base 10) of magnetic field strength, $\log{|\bi{B}|}$ at $z_{\mathrm{ob}} = 600 $ $ \mu\mathrm{ m}$ due to a TI-metamaterial layer structure generated by a 200-elementary-charge point electric source at $d_1 = 2 $ $ \mu\mathrm{ m}$ when TI is with $d_2 = 600\, \mathrm{ }\mu$m, $\theta_2 = \pi$, $\epsilon_2$ = 20 and $\mu_2$ = 1; and (a) $\epsilon_3 \in$ [-10.3, -9.9], $\mu_3 \in$ [-1.2, -0.8]; (b) $\epsilon_3 \in$ [-17.4, -17.0], $\mu_3 \in$ [-1.2, -0.8], with a step of 0.001.} \label{Fig:mu2_1_finemesh}
\end{figure}

To perform a detailed investigation on the effects of the metamaterial substrate properties on the induced magnetic field in free space, we chose the TI with $d_2 = 600 $ $ \mu\mathrm{ m}$, $\theta_2 = \pi$, $\epsilon_2$ = 20 and $\mu_2$ = 1. We then performed a fine sweep of the relative permittivity, $\epsilon_3$ and permeability, $\mu_3$ with a step of 0.001. From Fig.~\ref{Fig:mu2_1} (c), two sets of regions were identified and studied: (a) $\epsilon_3 \in$ [-10.3, -9.9], $\mu_3 \in$ [-1.2, -0.8] and (b) $\epsilon_3 \in$ [-17.4, -17.0], $\mu_3 \in$ [-1.2, -0.8]. The corresponding logarithm of magnetic field strength, $\log{|\bi{B}|}$ at $z_{\mathrm{ob}} = 600 $ $ \mu\mathrm{ m}$, generated by a 200-elementary-charge point electric source at $d_1 =2 $ $ \mu\mathrm{ m}$, is shown in Fig.~\ref{Fig:mu2_1_finemesh}. In Fig.~\ref{Fig:mu2_1_finemesh} (a) with $\epsilon_3 \in$ [-10.3, -9.9], $\mu_3 \in$ [-1.2, -0.8], the magnetic field strength in the free space at $z_{\mathrm{ob}} = 600 $ $ \mu\mathrm{ m}$ exhibits a symmetric pattern with distinct peaks and valleys around $\epsilon_3 = -10.092$ and $\mu_3 = -1$, with a maximum at $\epsilon_3 = -10.08$ $\mu_3 = -0.999$ which the induced magnetic field strength $|\bi{B}| = 249.2$ $\mu$T. In Fig.~\ref{Fig:mu2_1_finemesh} (b) with $\epsilon_3 \in$ [-17.4, -17.0], $\mu_3 \in$ [-1.2, -0.8], a similar pattern is observed. 

\begin{figure}[t]
\centering{}
\subfloat[]{ \includegraphics[width=0.45\textwidth]{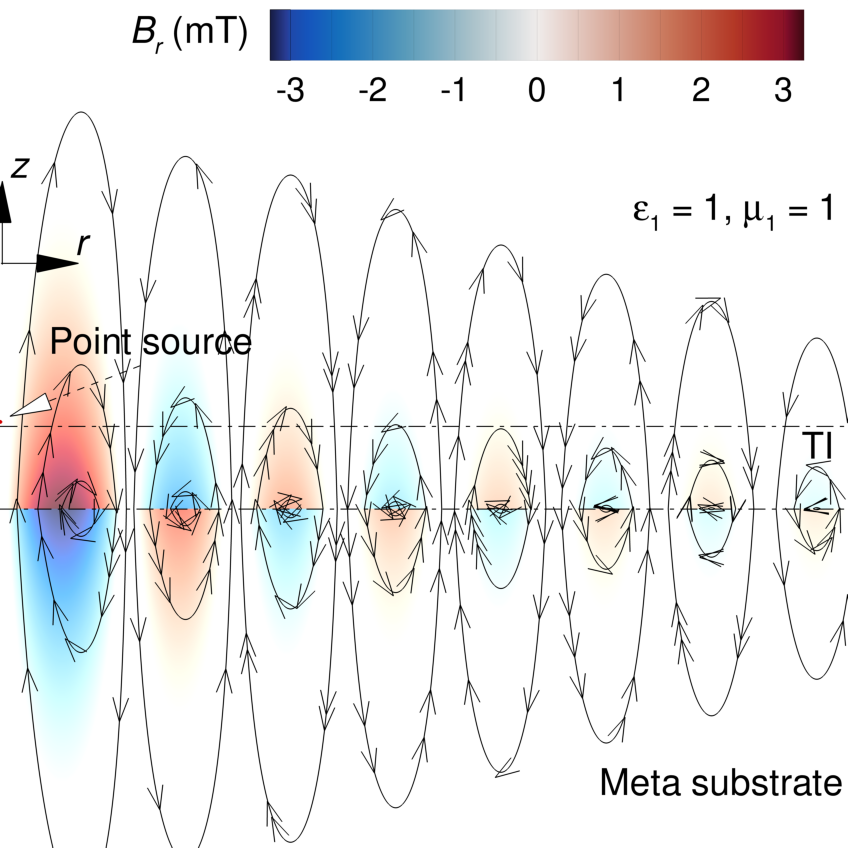}}  \quad
\subfloat[]{ \includegraphics[width=0.45\textwidth]{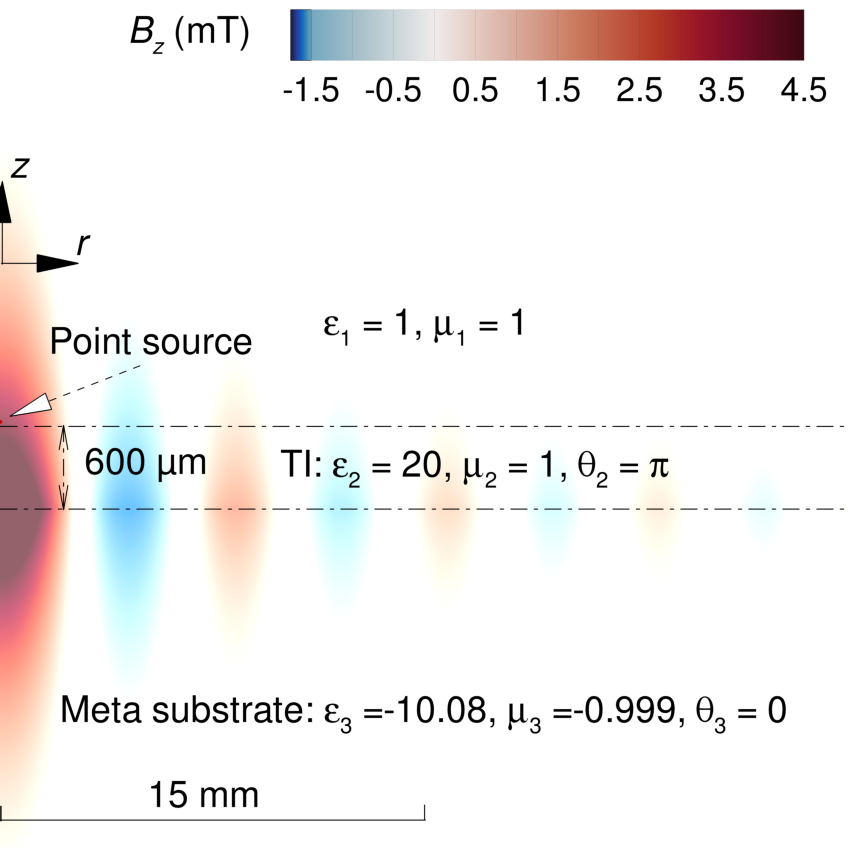}} 
\caption{The distributions of the magnetic field components in the free space, TI and metamaterial substrate, generated by a a 200-elementary-charge point electric source at $d_1 =  2 $ $ \mu\mathrm{ m}$ above TI with $d_2 =  600 $ $ \mu\mathrm{ m}$, $\theta_2 = \pi$, $\epsilon_2$ = 20 and $\mu_2$ = 1 on top of a metamaterial substrate with $\epsilon_3 = -10.08$ and $\mu_3 = -0.999$: (a) $r$ component and (b) $z$ component.} \label{Fig:BxBzfield}
\end{figure}
For the case when the metamaterial's permittivity and permeability are $\epsilon_3 = -10.08$ and $\mu_3 = -0.999$, respectively, the $r$ and $z$ components of the induced magnetic field by a 200-elementary-charge point electric source at $d_1 = 2 $ $ \mu\mathrm{ m}$ in free space, inside TI, and within the metamaterial substrate are shown in Fig.~\ref{Fig:BxBzfield}, in which the properties of the TI are $d_2 = 600  $ $ \mu\mathrm{ m}$, $\theta_2 = \pi$, $\epsilon_2$ = 20 and $\mu_2$ = 1, respectively. It is shown clearly that in this case, the magnetic field is significantly enhanced at the interface between the TI layer and metamaterial substrate, which radiates into free space. Along the $r$ direction, the $z$ component of the induced magnetic field oscillates with damping, which is consistent with the field driven by a point source, and its maximum occurs at the axis of the symmetry where the point source is located. The $r$ component of the magnetic field behaves similarly to the $z$ component, but its oscillation is out of phase with respect to the $z$ component. Along the $z$ direction, the $z$ component is continuous across the TI and metamaterial interface due to the continuity of the normal component of the magnetic field, while the $r$ component is in the opposite direction on the two sides of the interface due to the differences in the relative permittivity and permeability of the TI and metamaterial, in particular, the negative relative permittivity and permeability of the metamaterial.

\begin{figure}[t]
\centering{}
\subfloat[$\epsilon_{3} =-10.08$, $\mu_{3} = -0.999$]{ \includegraphics[width=0.45\textwidth]{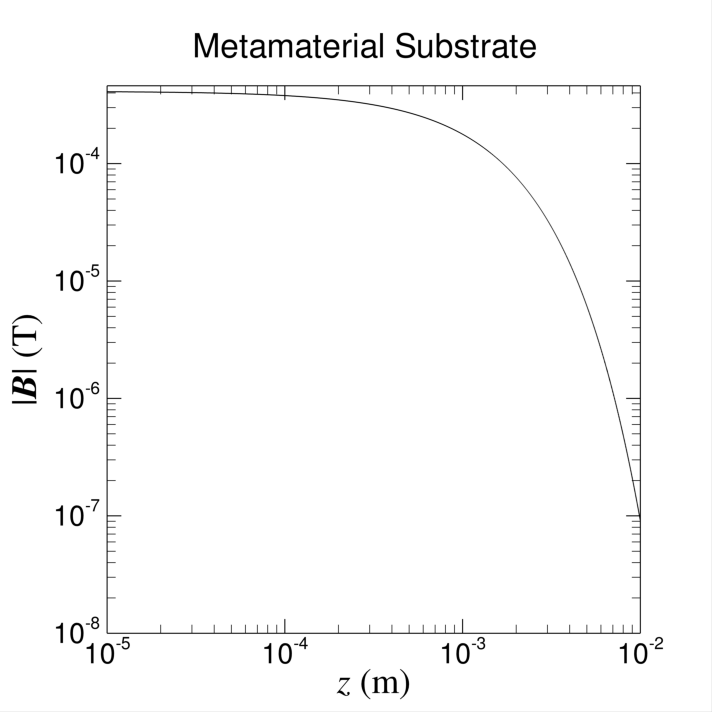}}  \quad
\subfloat[$\epsilon_{3} = 10.08$, $\mu_{3} = 1.000$]{ \includegraphics[width=0.45\textwidth]{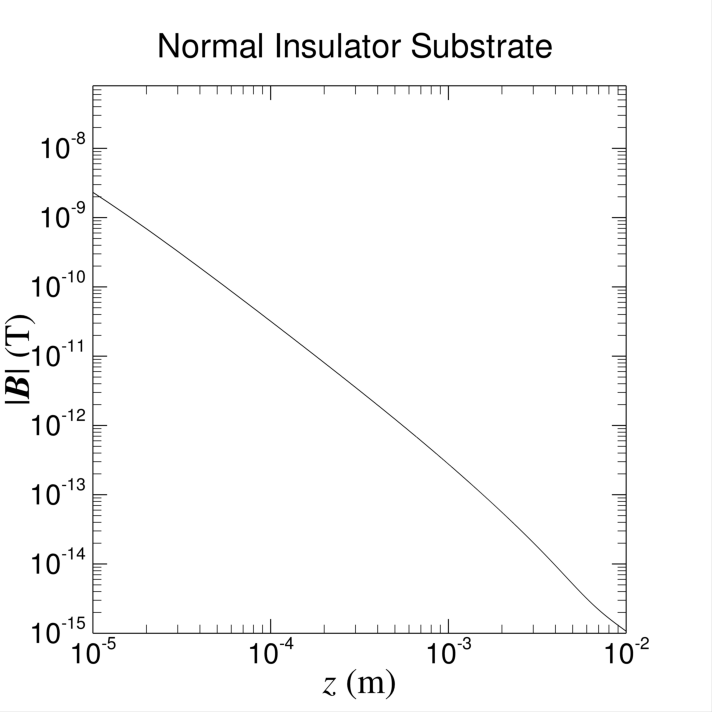}} 
\caption{The magnetic field strength, $|\bi{B}|$ at different $z$ when $r=0$, generated by a 200-elementary-charge point electric source at $d_1 = 2 \mathrm{ }\mu$m a TI with $d_2 = 600 \mathrm{ }\mu$m, $\theta_2 = \pi$, $\epsilon_2$=20, $\mu_2$=1 on: (a) metamaterial substrate with $\epsilon_{3} =-10.08$, $\mu_{3} = -0.999$, (b) normal insulator substrate $\epsilon_{3} = 10.08$, $\mu_{3} = 1.000$. } \label{Fig:mu2_1_line01}
\end{figure}

The magnetic field in free space decreases exponentially as the observation location moves away from the TI surface, as shown in Fig.~\ref{Fig:mu2_1_line01} (a). This gives us a relatively large region in free space with relatively high magnetic field. On the contrary, for the substrate as a normal insulator with almost the same magnitude of the relative permittivity and permeability, $\epsilon_3 = 10.08$ and $\mu_3 = 1.000$, respectively, the magnetic field amplitude is very small even next to the TI surface, and it decays nearly quadratically as the observation location moves away from the TI surface as presented in Fig.~\ref{Fig:mu2_1_line01} (b).

\begin{figure}[t]
\centering{}
\subfloat[$\mu_3 = -0.985$]{ \includegraphics[width=0.45\textwidth]{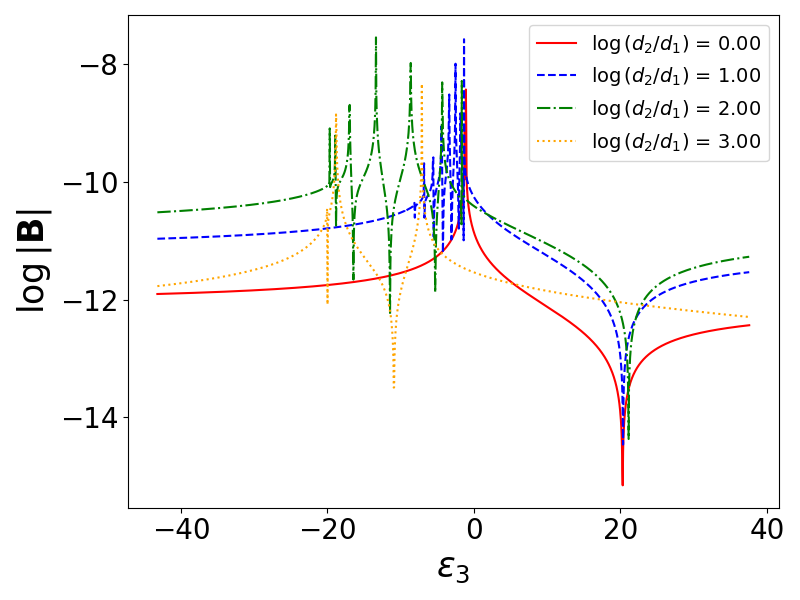}}  
\subfloat[$\epsilon_3 = -17.15$]{ \includegraphics[width=0.45\textwidth]{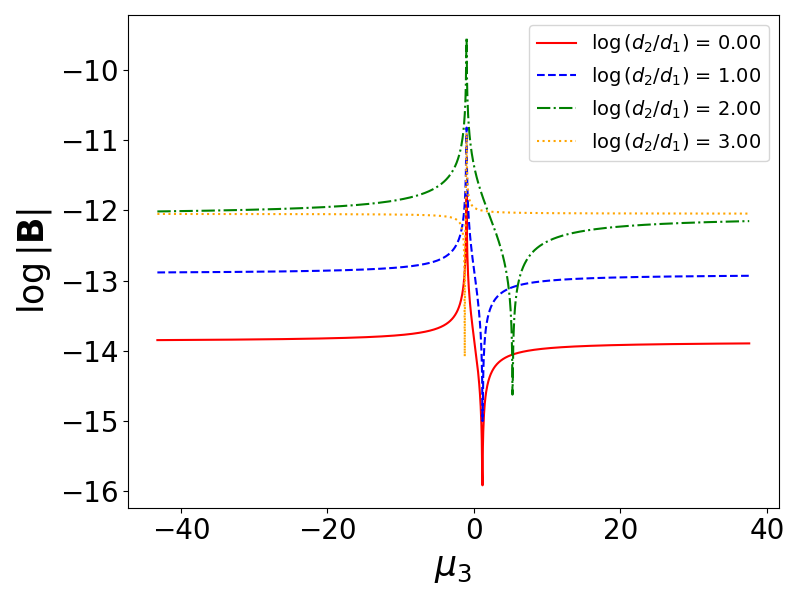}} \\
\subfloat[$\mu_3 = -0.985$]{ \includegraphics[width=0.48\textwidth]{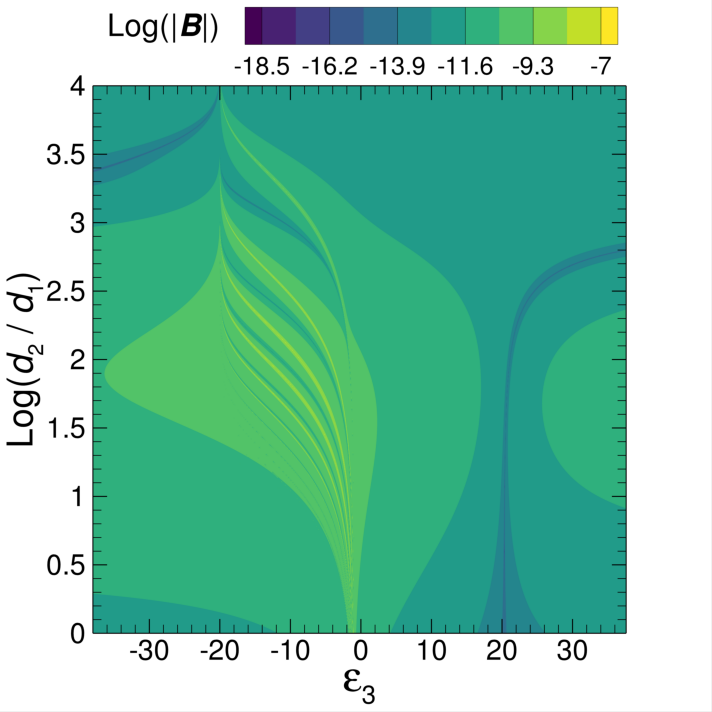}}  
\subfloat[$\epsilon_3 = -17.15$]{ \includegraphics[width=0.48\textwidth]{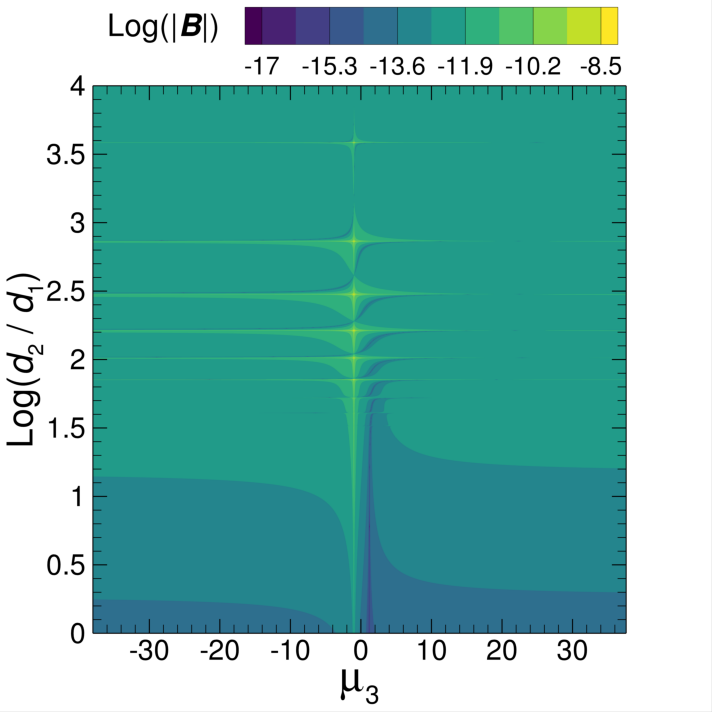}} 
\caption{The logarithm (base 10) of magnetic field strength, $\log{|\bi{B}|}$ at $z_{\mathrm{ob}} = 600 $ $ \mu\mathrm{ m}$, generated by a TI with different thickness when interfaced due to a 200-elementary-charge point electric source at $d_1 = 2 $ $ \mu\mathrm{ m}$ with various substrates when (a, c) its relative permittivity $\epsilon_3$ ranges from -35 to 35 and and (b, d) its permeability $\mu_3$ ranges from from -35 to 35. The bright peak features indicate conditions for maximum magnetic field enhancement, serving as a crucial guide for substrate selection aimed at boosting magnetic responses in TI applications. The number, position and sharpness of these peaks changes significantly due to the thickness of TI and substrate's relative permittivity and permeability.} \label{Fig:d2_2}
\end{figure}

We further investigate the effects of TI layer thickness on the induced magnetic field. Fig.~\ref{Fig:d2_2} provides a detailed view of how the thickness of a TI substrate and the electrical and magnetic properties of the substrate impact the magnetic field that was measured at $z_{\mathrm{ob}} = 600 $ $ \mu\mathrm{ m}$ induced by the point electric source with 200 elementary charges located at $d_1 = 2 $ $ \mu\mathrm{ m}$. The line plots in Fig.~\ref{Fig:d2_2} (a) and (b) provide a clear, quantitative illustration of how the magnetic field strength varies with TI thickness for fixed substrate parameters. Fig.~\ref{Fig:d2_2} (a) shows the magnetic field in free space when the substrate permeability is fixed at $\mu_3 = -0.985$, revealing a pronounced resonant peak in the magnetic field as the TI thickness increases. Similarly, Fig. 7 (b) illustrates the change of the magnetic field in free space while varying the TI thickness at a representative substrate permittivity of $\epsilon_3 = -17.15$. This plot demonstrates that for a given TI thickness, there exists an optimal substrate permeability that maximizes the magnetic field response.

Two contour maps in Figs.~\ref{Fig:d2_2} (c) and (d) depict more details of the logarithmic magnitude of the magnetic field strength as a function of the TI thickness $d_2$ and either the substrate's relative permittivity $\epsilon_3$ or its permeability $\mu_3$. Fig.~\ref{Fig:d2_2} (c) shows that the magnetic field strength has a complex dependency on the substrate's permittivity and TI thickness, with the permeability fixed at $\mu_3$ = -0.985. There are distinct regions where the magnetic field strength strength peaks, visible as bright lines. These peak lines indicate that at certain substrate permittivities, a resonant-like condition is achieved, which significantly enhances the magnetic response. When the thickness of TI is thin around $d_2/d_1 = 1$, the most significant enhancement of the magnetic field due to a point electric source occurs approximate at $\epsilon_3 = -1.5$. This effect becomes more pronounced and the peak areas broader as the TI layer becomes thicker. Fig.~\ref{Fig:d2_2} (d) maintains the substrate's permittivity at $\epsilon_3 = -17.15$ and varies its permeability and TI thickness. Here, the peak magnetic response shifts in position, implying that the optimal permeability for enhancing the magnetic field also varies with the thickness of the TI layer. There is an indication of optimal bands of permeability around $\mu_3 = -1$, which yield the highest magnetic field strength, especially as the TI layer increases in thickness. 

Thicker TI layers not only enhance the magnetic field but also offer a broader tolerance for variations in the substrate's permittivity and permeability. This implies a potential for tuning the system's sensitivity and performance by adjusting the TI layer's thickness, which can be critical for designing devices and applications that rely on precise control over magnetic field interactions.

\section{Discussion}

\begin{figure}[!t]
\centering{}
\subfloat[$\kappa d_2 = 0.01$]{ \includegraphics[width=0.45\textwidth]{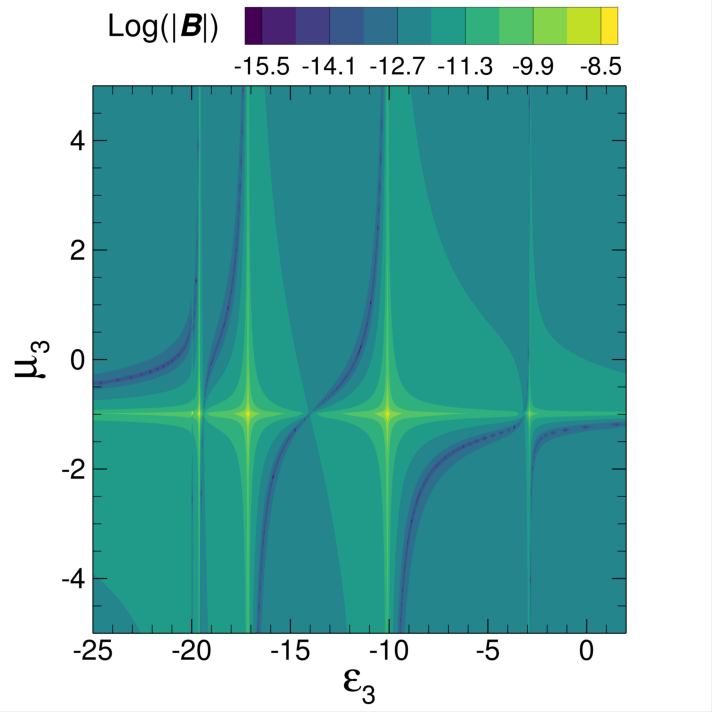}}  \quad
\subfloat[$\kappa d_2 = 0.1$]{ \includegraphics[width=0.45\textwidth]{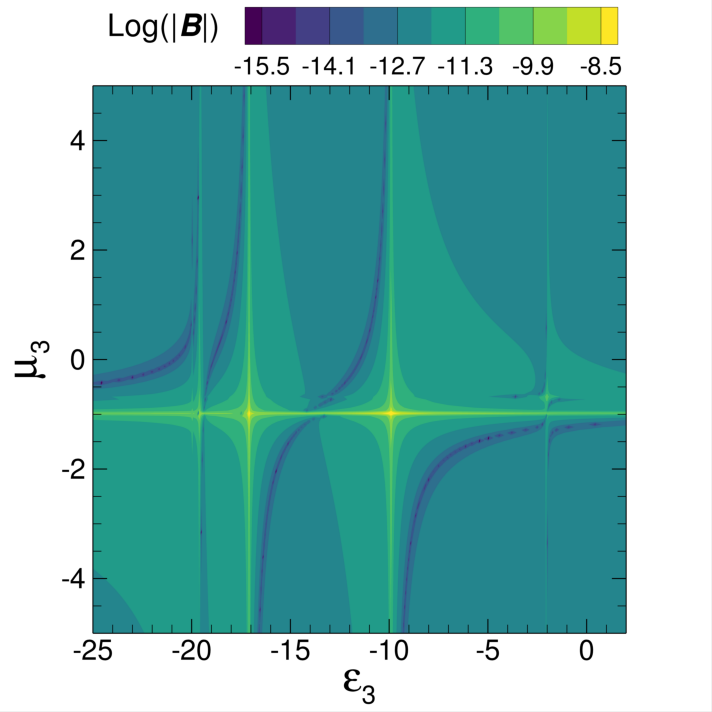}} \\
\subfloat[$\kappa d_2 = 1$]{ \includegraphics[width=0.45\textwidth]{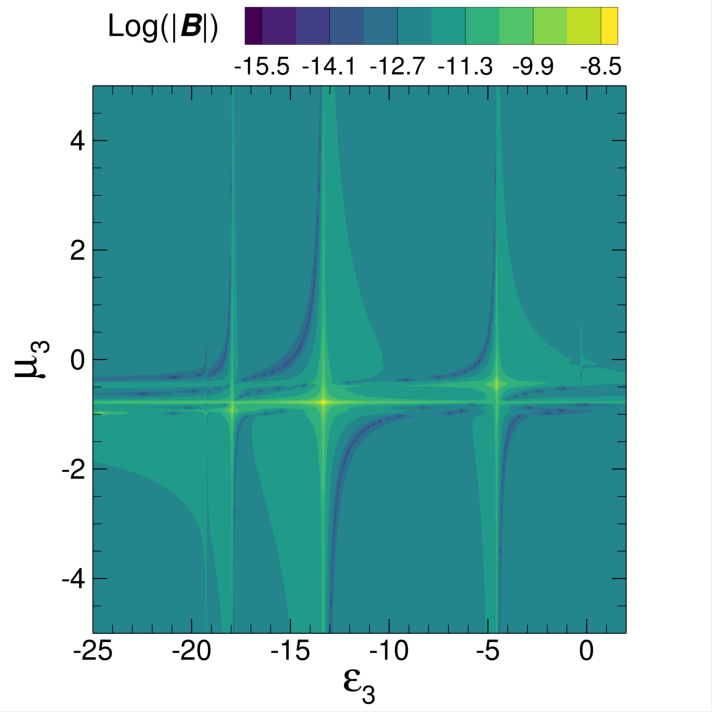}}  \quad
\subfloat[$\kappa d_2 = 5$]{ \includegraphics[width=0.45\textwidth]{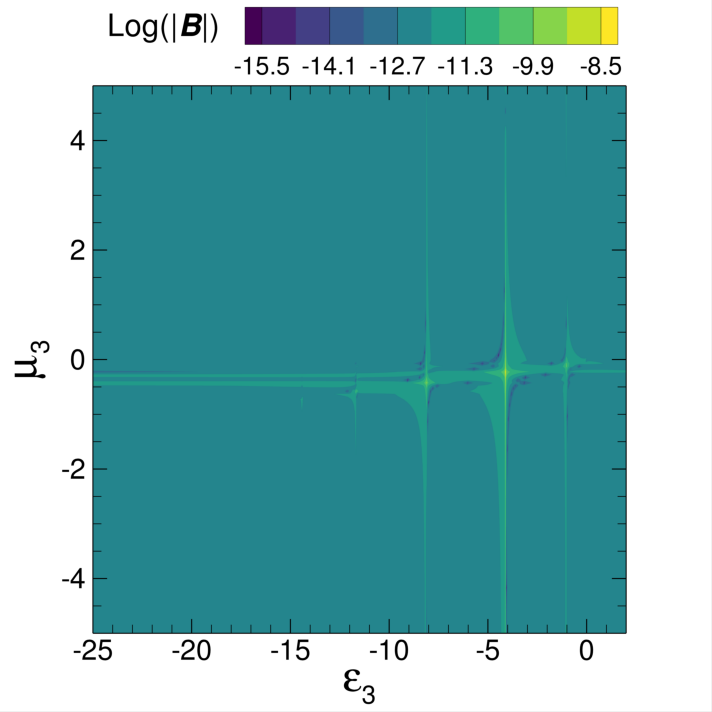}}
\caption{The logarithm (base 10) of magnetic field strength, $\log{|\bi{B}|}$ at $z_{\mathrm{ob}} = 600 $ $ \mu\mathrm{ m}$, generated by a TI with $d_2 = 600 $ $ \mu\mathrm{ m}$, $\epsilon_2$=20, $\mu_2=1$ and $\theta_2 = \pi$ when induced due to a 200-elementary-charge point electric source at $d_1 = 2$ $ \mu\mathrm{ m}$ on top of various metamaterial substrates with dissipated energy. When $\kappa d_2$ is small as in (a) and (b), the magnetic field enhancement in free space behaves nearly the same as when $\kappa \rightarrow 0$, as presented in Sec.~\ref{sec:results}. When $\kappa d_2 = 1$ as in (c), the magnetic field enhancement in free space due to the metamaterial is also very profound. When $\kappa d_2$ is much larger than 1, as in (d), the magnetic field response in free space is suppressed because the energy is absorbed quickly by the metamaterial.} \label{Fig:kappad2}
\end{figure}

In the previous section, we demonstrated that metamaterials can help enhance the magnetic field response of TIs, in particular when both the relative permittivity and permeability are negative. This can be achieved when these metamaterials are composed of active materials. 

There is some discussion that active, non-dissipated materials with negative permittivity and permeability could result in unstable physical systems \cite{Tretyakov2007}. To take energy loss into consideration, we modelled that the potentials for both the electric and magnetic fields in the metamaterial domain (domain 3) satisfy the Helmholtz equation, as
\begin{eqnarray}\label{eq:HelmholtzEq}
\nabla^2 \phi - \kappa^2 \phi= 0
\end{eqnarray}
where $\phi$ represents either $\phi^{3}_{E}$ or $\phi^{3}_{B}$ in the metamaterial domain, and $1/\kappa$ ($\kappa \gtrapprox 0$) is the characteristic length of energy loss, which is analogous to the skin depth of metallic materials or the Debye length of colloidal systems. When the value of $\kappa$ is low, the energy can be transmitted deeper into the metamaterial substrate relative to the case when the value of $\kappa$ is high.

In axis-symmetric cases, Eq. (\ref{eq:HelmholtzEq}) can also be solved effectively with the Hankel transform~\cite{Poularikas2009-dt}, which can be written as
\begin{eqnarray}\label{eq:Hankel0HelmEq}
\frac{\del^2 \Phi(\rho,z)}{\del z^2} - (\kappa^2+\rho^2) \Phi(\rho,z) = 0
\end{eqnarray}
for domain 3. As such, we need to rewrite Eq. (\ref{eq:Phi3}) as
\numparts
\label{eq:Phi3_Helm}
\begin{eqnarray}
\Phi_{E}^{3} &= C_{7} \frac{\exp{[\sqrt{\kappa^2+\rho^2} (z+d_{2})]}}{\sqrt{\kappa^2+\rho^2}}, \\
\Phi_{B}^{3} &= C_{8} \frac{\exp{[\sqrt{\kappa^2+\rho^2} (z+d_{2})]}}{\sqrt{\kappa^2+\rho^2}}.
\end{eqnarray}
\endnumparts
Correspondingly, to obtain coefficients $C_{7}$ and $C_{8}$ in Eq. (\ref{eq:Phi3_Helm}), the first two equations in Eq. (\ref{eq:C1C8_23}) need to be updated to 
\numparts
\label{eq:C1C8_23_Helm}
\begin{eqnarray}  
\fl \exp{[-\rho d_{2}]} \, C_{3} + C_{5} - \frac{\rho}{\sqrt{\kappa^2+\rho^2}} C_{7} = 0, \\
\fl \frac{1}{\mu_{2}} \Big\{\exp{[-\rho d_{2}]} \,  C_{4} + C_{6}  \Big\} + \frac{\theta_{2} \alpha}{\pi } \frac{1}{c} \Big\{\exp{[-\rho d_{2}]} \,  C_{3} + C_{5} \Big\}  - \frac{1}{\mu_{3}}\frac{\rho}{\sqrt{\kappa^2+\rho^2}} C_{8}  = 0.
\end{eqnarray}
\endnumparts

The results in Fig.~\ref{Fig:d2_2} indicate that the thickness of the TI has a clear effect on the magnetic field in free space. To demonstrate the interplay between the dissipated energy depth in the metamaterial substrate and the thickness of the TI, in Fig.~\ref{Fig:kappad2}, we plot the logarithmic magnitude of the magnetic field, $\log{|\bi{B}|}$, at $z_{\mathrm{ob}} = 600\, \mu\mathrm{m}$, generated by an electric point source with 200 elementary charges at $d_1 = 2\,\mu\mathrm{m}$, for different values of $\kappa d_2$. The thickness of TI is set to be $d_2 = 600 $ $ \mu\mathrm{ m}$, the relative permittivity $\epsilon_2$=20, the relative permeability $\mu_2=1$ and the topological phase $\theta_2 = \pi$. As we can see from Fig.~\ref{Fig:kappad2} (a) and (b), when $\kappa d_2$ is small---indicating the dissipated energy depth in the metamaterial substrate is much larger than the TI thickness---the magnetic field enhancement in free space behaves nearly the same as when $\kappa \rightarrow 0$, as presented in Sec.~\ref{sec:results}. When $\kappa d_2 = 1$, which means the energy loss depth in the metamaterial substrate is the same as the TI thickness, the magnetic field enhancement in free space due to the metamaterial is also very profound, as shown in Fig.~\ref{Fig:kappad2} (c). When $\kappa d_2$ is much larger than 1, the magnetic field response in free space is suppressed because the energy is absorbed quickly by the metamaterial.

\section{Conclusions}
This study demonstrates that by leveraging TI-metamaterial structures, the magnetic field induced by an electric point source in a TI can be significantly enhanced. Through detailed simulations and analysis, it was found that by carefully selecting the permittivity and permeability of the active metamaterial substrate, the induced magnetic field at the interface between the TI layer and the active metamaterial can be significantly enhanced, such as from a few pT to a few $\mu$T. This enhancement extends into free space, providing a larger region with a high magnetic field, which is crucial for exploring the novel magnetic field response of TIs. 

Our findings not only validate the effectiveness of the TI-metamaterial approach and serve as a benchmark for future numerical simulations, but also pave the way for future investigations. In upcoming work, we plan to explore more complex hybrid structures, including multilayer configurations and magnetized TIs~\cite{Lei2020, Liu2021}, to further elucidate the interplay between layered architectures and topological phenomena. Additionally, potential experimental validations can be achieved using techniques such as NV-center magnetometry~\cite{APinto2025}, which offers high spatial resolution and sensitivity for mapping the enhanced magnetic fields. Furthermore, our current study assumes inversion symmetry, which simplifies the analysis; future studies will address the effects of inversion symmetry breaking, such as the introduction of Rashba-type spin–orbit coupling, that may lead to new electromagnetic phenomena.

The practical implications of this research are substantial, with prospective applications in high-sensitivity magnetic sensors, advanced spintronic devices, tunable metamaterials for electromagnetic interference shielding, and novel platforms for quantum information processing. Overall, this work provides a promising pathway for both advancing the fundamental understanding of topological electromagnetic properties and developing innovative technologies based on these unique materials. 

\section*{Acknowledgments}
The authors acknowledge support from the Air Force Office of Scientific Research (AFOSR) FA2386-21-1-4125 for this work. This research was partially undertaken with the assistance of computing resources from RACE (RMIT AWS Cloud Supercomputing) and partially undertaken with the assistance of resources from the National Computational Infrastructure (NCI Australia), an NCRIS enabled capability supported by the Australian Government.

\appendix
\section{\label{sec:A} Coefficient $C_2$ in Eqs.~(\ref{eq:C1C8_12}) and (\ref{eq:C1C8_23})}

\begin{figure}[!t]
\centering{}
\subfloat[]{ \includegraphics[width=0.45\textwidth]{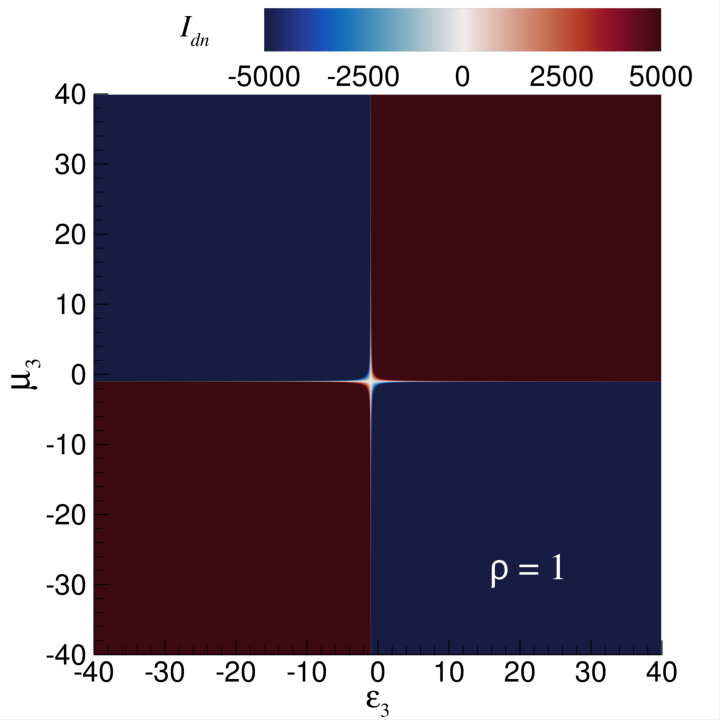}}  \quad
\subfloat[]{ \includegraphics[width=0.45\textwidth]{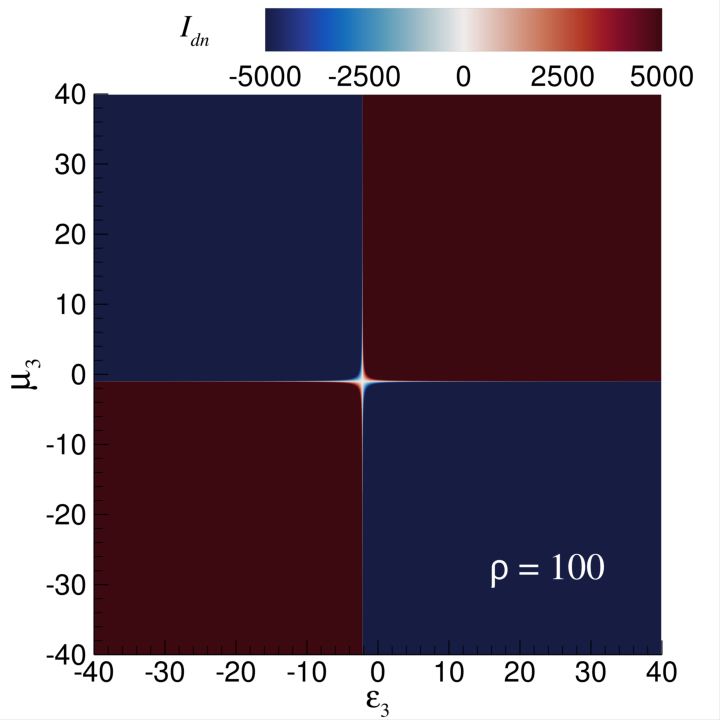}} \\
\subfloat[]{ \includegraphics[width=0.45\textwidth]{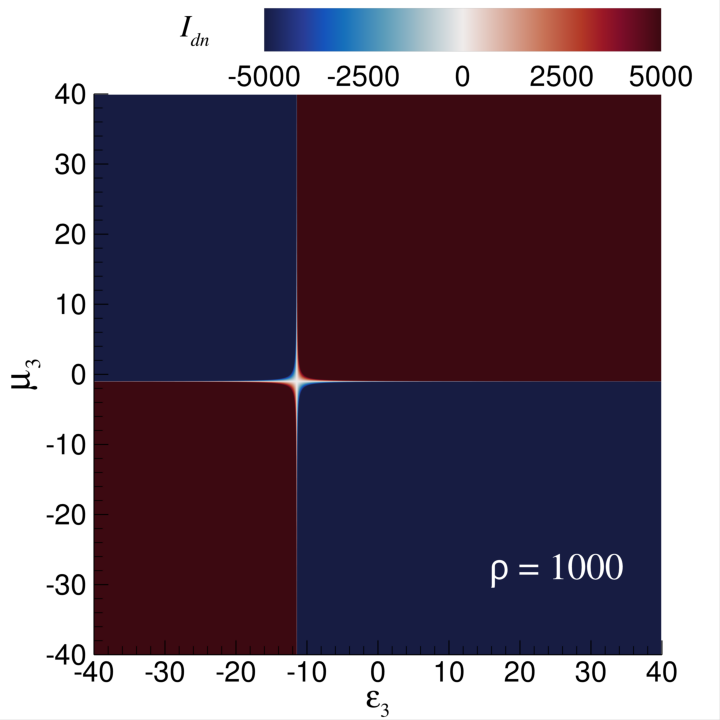}}  \quad
\subfloat[]{ \includegraphics[width=0.45\textwidth]{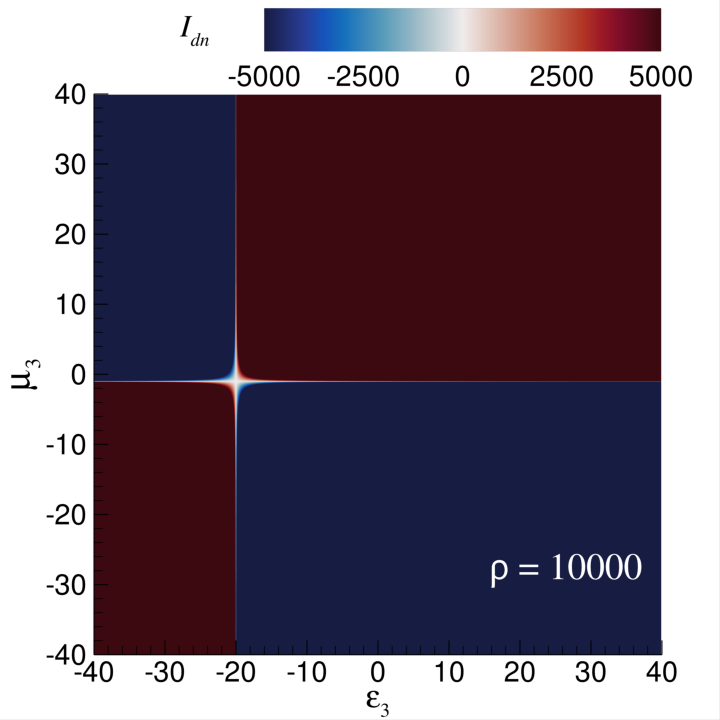}}
\caption{The variations of $I_{dn}$ (arbitrary unit) when the substrate relative permittivity $\epsilon_3$ and permeability $\mu_3$ sweep from -40 to -0.1 with the properties of the TI layer being $d_2 = 600 $ $ \mu\mathrm{ m}$, permeability $\mu_2=1$, relative permittivity $\epsilon_2$=20 and phase $\theta_2 = \pi$. Each sub-figure corresponds to a distinct value of Hankel transform variable $\rho$. The bright peak features indicate when $I_{dn}$ is close to zero.} \label{Fig:B_denom}
\end{figure}

\begin{equation}
C_2 = \frac{Q \alpha \theta_2 \mu_1 \mu_2}{c} \times \frac{I_{1} + I_{2} + I_{3} } {I_{\text{dn}}} 
\end{equation}
in which:
\begin{equation}
I_{1} = e^{-\rho(d_1 + z)  } \left[\alpha^2 \theta_2^2 \mu_2 \mu_3 + \pi^2 (\epsilon_2 + \epsilon_3) (\mu_2 + \mu_3) \right];
\end{equation}
\begin{equation}
I_{2} = e^{-\rho(d_1 + 2 d_2 + z)  } \left[ -2 \alpha^2 \theta_2^2 \mu_2 \mu_3 - 2 \pi^2 (\epsilon_3 \mu_2 + \epsilon_2 \mu_3)\right];
\end{equation}
\begin{equation}
I_{3} = e^{-\rho(d_1 + 4 d_2 + z)  } \left[ -\pi^2 (\epsilon_2 - \epsilon_3) (\mu_2 - \mu_3) + \alpha^2 \theta_2^2 \mu_2 \mu_3 \right] ;
\end{equation}
and 
\begin{equation}
    I_{\text{dn}} =  I_{4} + I_{5} + I_{6}
\end{equation}
with
\begin{equation}
    I_{4} = \left[\alpha^2 \theta_2^2 \mu_1 \mu_2 + \pi^2 (\epsilon_1 + \epsilon_2) (\mu_1 + \mu_2)\right] \left[\alpha^2 \theta_2^2 \mu_2 \mu_3 + \pi^2 (\epsilon_2 + \epsilon_3) (\mu_2 + \mu_3)\right],
\end{equation}
\begin{align}
    I_{5} = e^{-2\rho d_2  } \Big\{
    &-2 \alpha^4 \theta_2^4 \mu_1 \mu_2^2 \mu_3 - 
 2 \pi^2 \alpha^2 \theta_2^2 \mu_2 (\epsilon_3 \mu_1 \mu_2 + 
    2 \epsilon_2 \mu_1 \mu_3 + \epsilon_1 \mu_2 \mu_3)  \\ \nonumber
    & + 
 2 \pi^4 \left[\epsilon_1 \epsilon_2 \mu_2 (\mu_1 + \mu_3) + \epsilon_2 \epsilon_3 \mu_2 (\mu_1 + \mu_3) - \epsilon_2^2 (\mu_2^2 + \mu_1 \mu_3) - \epsilon_1 \epsilon_3 (\mu_2^2 +\mu_1\mu_3)\right] \Big\},
\end{align}
and
\begin{equation}
    I_{6} = e^{-4\rho d_2  } \left[\pi^2 (\epsilon_1 - \epsilon_2) (\mu_1 - \mu_2) - \alpha^2 \theta_2^2 \mu_1 \mu_2\right] \left[\pi^2 (\epsilon_2 - \epsilon_3) (\mu_2 - \mu_3) - \alpha^2 \theta_2^2 \mu_2 \mu_3\right].
\end{equation}

\section*{References}
\newcommand{\newblock}{}
\bibliographystyle{unsrt}
\bibliography{references.bib}

\end{document}